%% file: main_arxiv.tex
\documentclass[times,twocolumn,final]{elsarticle}

\usepackage{medima}
\usepackage{framed,multirow}
\usepackage{amsmath}
\usepackage{caption}
\usepackage{subfigure}
\usepackage{booktabs}

\usepackage{amssymb}
\usepackage{latexsym}

\usepackage{url}
\usepackage{xcolor}

\usepackage{hyperref}

\definecolor{newcolor}{rgb}{.8,.349,.1}

\journal{XXXX}

\begin{document}

\verso{Yihao Luo, Dario Sesia \textit{et~al.}}

\begin{frontmatter}

\title {
Explicit Differentiable Slicing and Global Deformation for Cardiac Mesh Reconstruction\tnoteref{tnote1}}

\tnotetext[tnote1]{Official Implementation: \hyperref[https://github.com/Luo-Yihao/GHDHeart]{https://github.com/Luo-Yihao/GHDHeart}}


\author[1]{Yihao \snm{Luo}\corref{cor1}\fnref{fn1}}
\cortext[cor1]{Corresponding authors}
\ead{y.luo23@imperial.ac.uk}
\author[8,7]{Dario \snm{Sesia}\fnref{fn1}}
\ead{d.sesia22@imperial.ac.uk}
\fntext[fn1]{The two authors contribute equally.}
\author[1,3,4]{Fanwen \snm{Wang}}
\author[1,3,4]{Yinzhe \snm{Wu}}
\author[1]{Wenhao \snm{Ding}}
\author[1,3,4]{Jiahao \snm{Huang}}
\author[1,3]{Fadong \snm{Shi}}
\author[2,6]{Anoop \snm{Shah}}
\author[2,7]{Amit \snm{Kaura}}
\author[2,7]{Jamil \snm{Mayet}}
\author[1,7,3,4]{Guang \snm{Yang}\corref{cor1}\fnref{fn2}}
\ead{g.yang@imperial.ac.uk}
\author[1]{ChoonHwai \snm{Yap}\corref{cor1}\fnref{fn2}}
\ead{c.yap@imperial.ac.uk}
\fntext[fn2]{Co-last Senior Authors: Guang Yang and Choon Hwai Yap}

\address[1]{Department of Bioengineering, Imperial College London, London, UK}
\address[8]{Department of Medicine, Imperial College London, London, UK}
\address[2]{Department of Cardiology, Imperial College Healthcare NHS Trust, London, UK}
\address[7]{National Heart \& Lung Institute, Imperial College London, London, UK}
\address[3]{Imperial-X, Imperial College London, London, UK }
\address[4]{Cardiovascular Research Centre, Royal Brompton Hospital, London, UK}
\address[5]{British Heart Foundation Centre of Research Excellence, Imperial College London, UK}
\address[6]{London School of Hygiene and Tropical Medicine, London, UK}

\received{XXXX}
\finalform{XXXX}
\accepted{XXXX}
\availableonline{XXXX}
\communicated{XXXX}

\begin{abstract}
Three-dimensional (3D) mesh reconstruction of the cardiac anatomy from medical images is useful for shape and motion measurements and biophysics simulations to facilitate the assessment of cardiac function and health. However, 3D medical images are often acquired as 2D slices that are sparsely sampled (e.g., large slice spacing) and noisy, and 3D mesh reconstruction on such data is a challenging task. Traditional voxel-based approaches utilize pre- and post-processing that compromises fidelity to images, while mesh-level deep learning approaches require large 3D mesh annotations that are difficult to get. Therefore, direct cross-domain supervision from 2D images to 3D meshes is a key technique for advancing 3D learning in medical imaging but it has not been well-developed. While there have been attempts to approximate the voxelization and slicing of meshes that are being optimized, there has not yet been a method for directly using 2D slices to supervise 3D mesh reconstruction in a differentiable manner. Here, we propose a novel explicit differentiable voxelization and slicing (DVS) algorithm allowing gradient backpropagation to a 3D mesh from its slices, which facilitates refined mesh optimization directly supervised by the losses defined on 2D images. Further, we propose an innovative framework for extracting patient-specific left ventricle (LV) meshes from medical images by coupling DVS with a graph harmonic deformation (GHD) mesh morphing descriptor of cardiac shape that naturally preserves mesh quality and smoothness during optimization. The proposed framework achieves state-of-the-art performance in cardiac mesh reconstruction tasks from densely sampled (CT) as well as sparsely sampled (MRI stack with few slices) images, outperforming alternatives, including marching cubes, statistical shape models, algorithms with vertex-based mesh morphing algorithms and alternative methods for image-supervision of mesh reconstruction. Experimental results demonstrate that our method achieves an overall Dice score of 90\% during a sparse fitting on multi-datasets. The proposed method can further quantify clinically useful parameters such as ejection fraction and global myocardial strains, closely matching the ground truth and outperforming the traditional voxel-based approach in sparse images.

\end{abstract}


\end{frontmatter}

\input{introduction/introduction}

\input{methods/methods}

\input{results/exp}

\input{discussion/discussion}




\bibliographystyle{model2-names.bst}\biboptions{authoryear}
\bibliography{bibs/bibliography}

\appendix

\end{document}

%% file: introduction/introduction.tex
\section{Introduction}
Cardiovascular diseases are one of the leading causes of mortality worldwide, accounting for around 17.9 million deaths annually \cite{roth2015demographic}. Computational reconstruction of the cardiac anatomy is useful for shape and motion measurements to determine heart function for diagnosis, treatment and prognosis \cite{brady2023myocardial}. They are also useful for physics-based modelling to understand cardiac physiology and predict intervention outcomes, and also for surgical planning \cite{hashim2006finite, green2024pre}. 
However, reconstructing patient-specific cardiac mesh is challenging because of imaging imperfection (insufficient resolution or excessive noise) and the variability of cardiac shape across individuals.

To date, most studies have focussed on the segmentation of voxelated medical images rather than a mesh-level reconstruction. When the final mesh result is desired, investigators typically use offline algorithms for isosurface extraction from segmentations and contours (\cite{fedele2021polygonal,villard2018surface}), via the marching cubes algorithm (\cite{lorensen1998marching}) and point cloud surface reconstruction(\cite{hoppe1992surface}). However, meshes reconstructed by these algorithms often suffer from staircase artifacts, uncontrollable topology, high polygon counts, and poor smoothness. These issues can lead to inaccuracies in representing the complex geometries of cardiac structures, and smoothing post-processing to address staircasing can deteriorate its fidelity to images. 
Further, these methods involve multi-step procedures without differentiability, which 
prevents the integration of this mesh reconstruction approach into deep learning frameworks for better optimization and learning.
In sparsely sampled data, it is even harder to apply marching cubes as this will lead to much more topological flaws, and pre-processing the spare images with interpolations will also deteriorate fidelity to images. 

A 3D Mesh, as a discretization of a non-Euclidean manifold (\cite{meyer2003discrete}), is hard to be directly optimized by deep learning frameworks until the recent development of graph convolutional networks (GCN) (\cite{wu2020comprehensive}). Some studies proposed deep learning-based methods to directly predict the mesh results (\cite{kong2020automating, kong2021deep, beetz2022interpretable, meng2023deepmesh, hattori2021deep}). These methods typically combine a convolutional neural network (CNN) for image feature extraction and a graph convolutional network (GCN) for mesh processing. 
However, these data-driven methods require large, high-quality training datasets and ground truths, which can be expensive or even impossible to obtain. 
Several works resorted to using marching cubes to obtain such training datasets, thus retaining limitations of marching cubes. 
Therefore, it is desirable to develop a differentiable algorithm to simulate the slicing of 3D objects, which allows the gradient of a loss function defined on a 2D image to be back-propagated to the mesh. This can be used to optimize the mesh directly with the image-level supervision, where the image-level annotations are much easier to obtain. To the best of our knowledge, there is no existing explicit differentiable algorithm for this purpose before this work, though some works have proposed alternative approximating methods like \cite{hattori2021deep, meng2023deepmesh}. Thus, the first contribution of this work is to propose a differentiable voxelization and slicing (DVS) algorithm for establishing direct, global supervision of image labels on the slices of a pending-to-optimize mesh without approximating errors.

Meanwhile, mesh-level learning for 3D cardiac reconstruction is usually achieved by optimizing a learnable vertex-wise deformation from a template mesh, which can cause oscillations and self-intersections unless additional regularization terms are introduced.
Inspired by the graph Fourier analysis \cite{sandryhaila2013discrete}, we propose a global mesh deformation algorithm based on graph harmonics (GHD), which decomposes the deformations of a mesh as surface Fourier waves to preserve geometric information at different levels. Representing the deformation as a global function on the mesh, rather than a local function on each vertex, can preserve mesh smoothness and quality during the optimization process, thereby overcoming the limitations of current vertex-wise deformation techniques and providing a more robust framework for cardiac mesh reconstruction. Compared to the existing mesh morphing methods and statistical shape models (SSM), such as \cite{upendra2021cnn, pak2021weakly, bai2015bi,carminati2018statistical}, our GHD algorithm tends to take the balance between the dimensional reduction and the degree of freedom and avoid the dependency on the training data, which is more flexible and efficient for the mesh reconstruction.

By combining DVS and GHD, we propose a novel differentiable framework for cardiac mesh reconstruction adaptive to dense or sparse segmentation slices. This framework is flexible and can incorporate various differentiable operators as auxiliary constraints. It does not require pre- and post-processing, smoothness regularization, or prior training. It achieves SOTA performance on different datasets in both CT and MRI.

\begin{figure*}[!t]
\centering
\includegraphics[width =1\textwidth ]{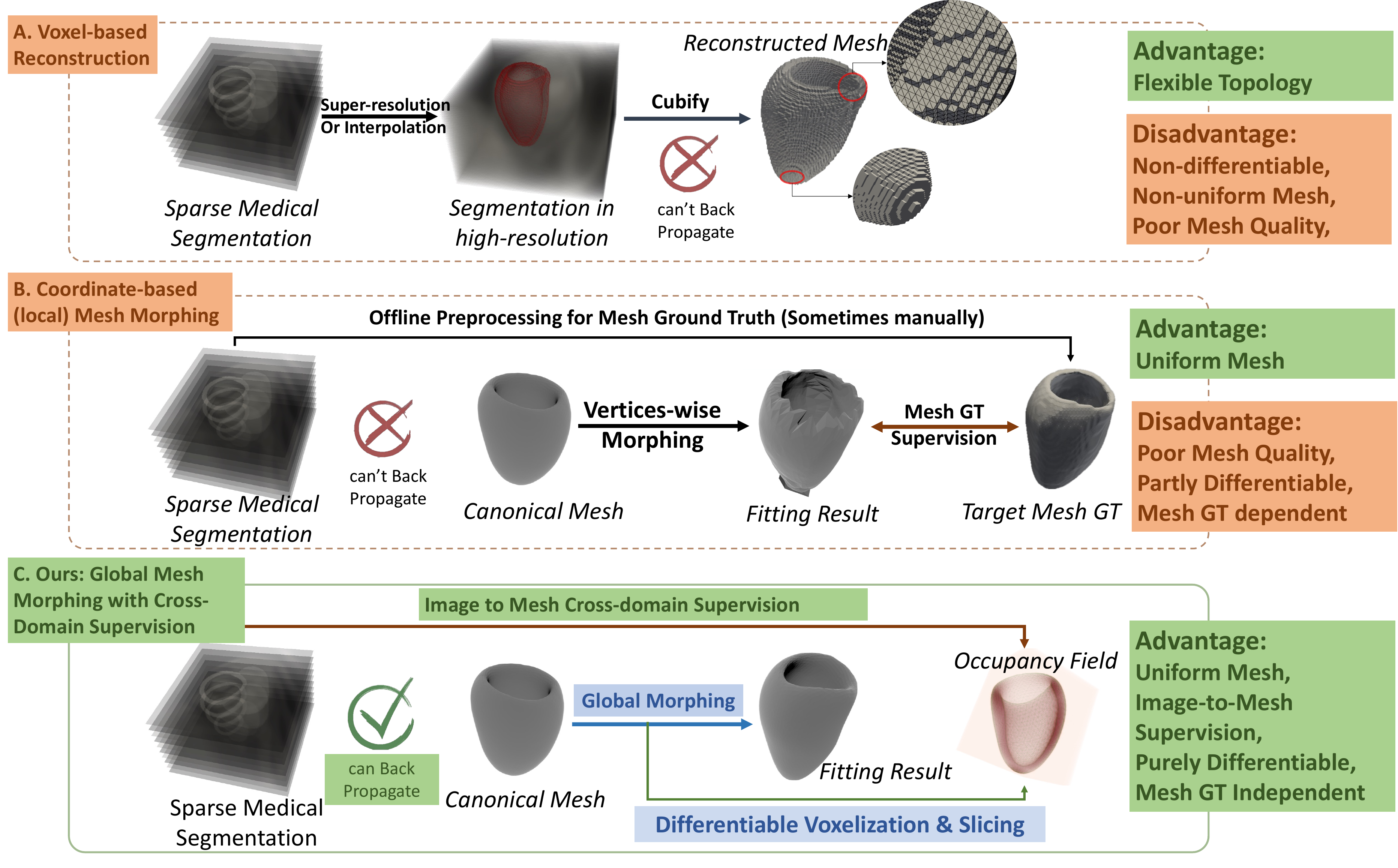}
\captionsetup{justification=raggedright, margin=0.1cm}
\caption{\small The Motivation and Novelty of our Proposed Method. Block A depicts the image-level learning and post-processing for 3D mesh reconstruction. Block B shows previous mesh-level optimization approaches with vertex-wise deformation and mesh ground truth (GT) supervision. Block C introduces our proposed method, which incorporates differentiable mesh operations and global mesh deformation based on graph harmonics (GHD), enabling direct supervision between meshes and images. Green marks indicate the advantages of methods,  red marks indicate the weaknesses. The blue arrows highlight the key technical contributions of our work: differentiable voxelization, slicing (DVS) and the global deformation GHD.}
    \label{fig: novelty}
\end{figure*}

\subsection{Related Work}
In the following, we review the existing methods for cardiac mesh reconstruction and discuss their limitations.

\subsubsection{Voxel-Based Cardiac Mesh Reconstruction}
Traditionally, 3D mesh-based heart reconstruction from labeled images is conducted via the marching cubes algorithm (\cite{Mantilla2023, Lechelek2022}), coupled with smoothing post-processing to remove the staircase texture. An example of this staircase texture can be seen in Fig. \ref{fig: novelty}. However, the marching cubes algorithm has several limitations. Firstly, it relies on the assumption of a uniform voxel grid, which can lead to inaccuracies in representing the complex geometries of cardiac structures, smoothing post-processing, and deteriorating its fidelity to images. Additionally, the algorithm often produces meshes with many polygons, necessitating further simplification and optimization steps to make the meshes usable for computational modeling, which can introduce additional errors and distortions.

Moreover, the marching cubes algorithm inherently lacks differentiation capabilities, which limits its integration with modern machine learning frameworks that require differentiable operations for optimization and learning. This non-differentiability hinders the development of end-to-end deep learning models for cardiac mesh reconstruction, where back-propagation through the entire pipeline, including the reconstruction step, is essential. More recently, some works have proposed differentiable versions of the marching cubes algorithm, such as \cite{shen2023flexicubes,liao2018deep}. These methods show promise in enabling the integration of mesh reconstruction into deep learning frameworks, facilitating end-to-end optimization and learning. However, without cross-domain supervision, these methods still rely on 3D label for training, which prevents their applications in medical imaging tasks.

\subsubsection{Cardiac Statistical Shape Models (SSM)}
SSM is another approach for image-based mesh reconstruction. It is traditionally derived from principal component analysis (PCA)(\cite{frangi2002automatic,grosgeorge2013graph}) and proper orthogonal decomposition (POD) ((\cite{fresca2021pod})). These methods typically use training data to learn a low-dimensional parametric shape model that captures the variability of the cardiac shape across the population. Subsequently, a weighted sum of the shape modes is fitted to the images for the mesh reconstruction. For example, \cite{bai2015bi} developed a PCA-based statistical shape model from an atlas of over 1000 MR images. \cite{carminati2018statistical} developed a similar model with echocardiography scans of 435 patients to guide segmentations in other scan modalities. 
Besides, deep learning-based SSMs have been proposed, such as \cite{chen2020anatomy}, which uses a variational autoencoder (VAE) to learn the shape modes from a large dataset of cardiac MRI images. These methods have shown high accuracy in reconstructing cardiac meshes from images.
However, this approach still utilizes marching cubes to derive training mesh data and retain the limitations of marching cubes. Further, training meshes must all have a uniform mesh structure with a fixed number of nodes, which can limit mesh quality and increase the data preparation burden. 
It would thus be advantageous to have a technique that does not require large data training, such as our GHD+DVS approach.

\subsubsection{Mesh-based Deep Learning Cardiac Reconstruction}
Deep learning approaches have shown great promise in cardiac mesh reconstruction. For instance, \cite{kong2021deep} developed a deep learning framework for direct whole-heart mesh reconstruction directly from 3D image data with promising accuracy. Their methods conventionally involve a segmentation module with a U-Net architecture (\cite{jia2019automatically, tong20183d}) that allows supervision using ground truth annotations, followed by a mesh generation module that predicts the mesh vertices from the segmented image with a graph convolutional network (\cite{wu2020comprehensive}). Another approach is mesh-based VAE models proposed by \cite{beetz2022interpretable} for non-linear dimensionality reduction and cardiac mesh generation, which also demonstrates high accuracy and physiological reconstructions. However, both approaches relies on manual preparation for mesh ground truths, which is time-consuming, and shares the accuracy limitations of traditional voxel-based methods. It would thus be advantageous to have a technique that does not require large data training, and avoid traditional voxel-based methods, such as our GHD+DVS approach.



Mesh morphing methods establish a mean template mesh of the heart, and the reconstruction process involves deforming the template mesh using local optimizations to match tissue boundaries on input images. For example, \cite{upendra2021cnn} proposed a method to warp a patient-specific ED volume mesh based on registration-based propagated surface meshes using a log barrier-based mesh warping (LBWARP) method. \cite{pak2021weakly} proposed a method that applies a registration-based deformation field to mesh vertices, supervised by the Chamfer distance (CD) between the warped mesh and the target mesh.

Mesh morphing methods can be sensitive to the template initialization, which may require complicated steps and manual efforts for mitigation. However, this approach can be differentiable, and if managed well, can produce smooth reconstructions with high image fidelity. Currently approaches, as above, utilizes a marching cube mesh database for training, this can be avoided if it is coupled with our proposed DVS. 

The advantages of deep learning approaches include their differentiability, which allows integration into end-to-end optimization frameworks, and their ability to learn complex, non-linear mappings from data, leading to high reconstruction accuracy. However, similar to statistical shape models, these approaches are data-driven and require large, high-quality training datasets and ground truths, which can be challenging to obtain.

\subsubsection{Approximating Cross-domain Supervision}
The typical way to facilitate image supervision of mesh reconstruction takes the form of a differentiable loss function describing the match between the mesh nodes and image voxels or image pixels along several image planes or slices, which can be optimized during the reconstruction process.

For example, the fitting of statistical shape modes or mesh morphing template meshes to images requires a differentiable module that minimizes a loss term describing the proximity of mesh nodes to image boundaries. This is typically achieved via distance-based measures between mesh nodes and image boundary voxels, such as the Chamfer distance. However, these are local loss functions that are likely to generate local minima, making convergence more challenging. Having a global loss function can minimize this effect and enable better convergence and accuracy.

The differentiable mesh-to-image rasterizing proposed by \cite{meng2023deepmesh} compares the image-level mask contour with probabilistic mesh intersections, considering only the boundary information. This approach, however, is again a local loss function with the above limitations. Further, the projection of only the mesh boundary to the rasterization plane can be noisy. Another approach, the approximately differentiable voxelization and slicing (ADVS) method proposed by \cite{hattori2021deep}, performs an initial offline voxelization of the canonical template, followed by voxelated image warping guided by iterative mesh fitting. This method also relies on a local distance-based loss. Moreover, supervisions are implemented at the image level, meaning that back-propagation does not directly relate to mesh-level optimization but stops at the initial voxelization mask.

Our proposed differentiable DVS algorithm utilizes an effective global loss function for the image supervision of mesh reconstruction. Unlike the deep learning approaches above \cite{meng2023deepmesh, hattori2021deep}, DVS can achieve direct supervision at the mesh level, rather than an indirect one at the image level, to facilitate back-propagation, to improve convergence and thus accuracy.


\subsubsection{Global Mesh Deformation}
A challenge faced by mesh morphing mesh reconstruction approaches is the smoothness and quality of generated meshes. Existing reconstruction techniques, which often rely on vertex-wise deformation, can cause oscillations and self-intersections unless additional regularization terms are introduced. Mesh quality is importance for biophysics simulations and for feature extraction in mesh-based deep learning, but is not well studied. Our proposed GHD approach can address these limitations.

Graph Fourier analysis is a cutting-edge topic within signal processing, particularly for data on non-Euclidean domains such as graphs and meshes. The Graph Fourier Transform (GFT) extends the classical Fourier transform to graphs, utilizing the eigenvalues and eigenvectors of the graph Laplacian matrix for frequency analysis. The graph Laplacian, defined as \(L = D - A\) (where \(D\) is the degree matrix and \(A\) is the adjacency matrix), is fundamental to GFT. The eigenvalues represent the graph frequencies, and the eigenvectors serve as the basis for the transform. A graph signal is a function defined on the graph's nodes, with each node associated with a signal value. The GFT projects a graph signal onto the eigenvector basis of the Laplacian, enabling analysis in the spectral domain. This approach has applications in various fields, including network analysis, image processing, and machine learning on graph-structured data \cite{shuman2013emerging}.

Mesh spectral processing extends these concepts to 3D meshes, which are crucial in computer graphics and geometric processing. A 3D mesh, typically represented by vertices and faces, can be analyzed using the mesh Laplacian, analogous to the graph Laplacian. The spectral decomposition of the mesh Laplacian facilitates operations like smoothing, compression, and shape analysis. Both graph Fourier and mesh spectral processing provide powerful tools for handling complex data structures, enabling more effective and efficient analysis and manipulation of signals on graphs and meshes. Key references in this field include foundational works by Shuman et al. on graph signal processing \cite{shuman2013emerging} and by Zhang et al. on spectral mesh processing \cite{zhang2010spectral}, which lay the groundwork for these transformative techniques.

Inspired by these works \cite{zhang2010spectral, rong2008spectral}, our GHD approach considers the deformations of a mesh as decomposed on spectral bases to preserve geometric information at different levels. Based on the graph spectral and GFT theory, we propose a novel optimization technique for generating patient-specific left ventricle (LV) meshes via global deformations. This method ensures the smoothness and quality of the generated meshes, thereby overcoming the limitations of current vertex-wise deformation techniques and providing a more robust framework for cardiac mesh reconstruction.

\subsection{Contributions} 
Our specific contributions can be summarized as follows:
\\
\\
\textbf{Differentiable Voxelization and Slicing (DVS) Algorithm.} We propose the DVS algorithm for establishing direct, global supervision of image labels on mesh reconstruction. Compared to local distance-wise supervision approaches and indirect supervision approaches, this improves convergence and accuracy, contributing to a more robust ability to reconstruct from sparsely sampled images. The DVS algorithm ensures that the back-propagation process directly relates to mesh-level optimization, facilitating more precise and reliable cardiac mesh reconstructions. This technical can be widely used in general medical 3D reconstruction tasks. 
\\
\\
\textbf{Graph Harmonics Deformation (GHD) Algorithm.} We introduce the novel GHD algorithm for generating patient-specific left ventricle (LV) meshes quickly by morphing from a canonical mesh, where displacements are described as surface Fourier waves. GHD naturally preserves high mesh quality and smoothness, enabling robust performance even in sparsely sampled medical images. This enhances the accuracy and efficiency of cardiac mesh reconstruction.
\\
\\
\textbf{Novel Differentiable Framework for Cardiac Mesh Reconstruction from 2D Slices.} By combining DVS and GHD, we propose a robust framework for differentiable cardiac mesh reconstruction adaptive to dense or sparse segmentation slices. This framework has the flexibility of incorporating various differentiable operators as auxiliary constraints, does not requiring pre-and post-processing, smoothness regularization, or prior training. It achieves SOTA performance on different datasets in both CT and MRI.


%% file: methods/methods.tex
\section{Theoretical Framework}

\subsection{Differentiable Voxelization \& Slicing (DVS)}

Voxelization and slicing are essential for image supervision on 3D cardiac mesh reconstructions. Traditional mesh-boolean-based algorithms (\cite{de2000computational}) are not differentiable and cannot be integrated into optimization frameworks such as neural networks. As a result, pre- and post-processing steps are often required to obtain 3D results from image-level training, complicating visualization and evaluation.

Previous work like \cite{meng2023deepmesh, joyce2022rapid} have attempted to address this by using probabilistic rasterization and image-level warping for implicit voxelization and slicing. However, these methods have limitations in efficiency and accuracy of information transfer between images and meshes.

Our proposed differentiable DVS algorithm can overcome these challenges, and provide direct supervision between meshes and images, facilitating mesh-level optimization in medical image deep learning.

Our method is inspired by classical physics field theory, where vector fields emanating from a source decay in strength according to the inverse square law with distance from the source, as seen in gravitational \cite{head2003gravity} and electric fields \cite{van2007electromagnetic}. In this context, every point within a cardiac mesh generates a vector field that influences its occupancy within the mesh. By morphing the mesh to maximize occupancy of all voxelated points designated within the mesh on the image, we can achieve an optimal fit to the image data.

Mathematically, the field strength \( \vec{E_q} \) at a point \( x \) from a single source \( q \) satisfies:

\begin{equation}\label{eq:field}
    \iint_{\partial \Omega} \left \langle \vec{n},\vec{E_q} \right \rangle ds = \iiint_\Omega \nabla \cdot \vec{E_q} dv = 4\pi \delta _q,
\end{equation}
\\
where \( \delta _q \) equals 1 if \( q \in \Omega \) or 0 if not, and \( \Omega \) is a 3D volume with \( \partial \Omega \) as its meshed surface. The abstract inverse quadratic field is defined as the unit field direction vector multiplied by the inverse quadratic field strength:

\begin{equation}
    E_q(x) = \frac{x-q}{\vert\vert x-q\vert\vert }\cdot\frac{1}{(x-q)^2} =  \frac{x-q}{\vert\vert x-q \vert\vert^3 }.
\end{equation}

The Gauss Theorem states that the surface integration of the vector flux along \( d\Omega \) is proportional to the total influence of the sources within the surface \cite{gauss1877theoria}. Conversely, this surface integration can determine the number of sources within \( d\Omega \) or the binary occupancy of a point source. This is known as the Winding Number Theorem in topology \cite{chillingworth1972winding}. 

Our DVS algorithm is formulated as the discrete version of Equation \eqref{eq:field}. To determine whether a point is within a mesh boundary, we construct its vector field, \( \vec{E} \), and calculate vectors at each vertex of the mesh, \( d\Omega \). Next, at every vertex, we seek the inner product of \( \vec{E} \) and the normal vector \( \vec{n} \), and sum them after weighting with surface area represented by the vertex area of dual faces \cite{meyer2003discrete}. The occupancy of the query source point, \( q \), is calculated by:

\begin{equation}\label{eq:occupancy}
     \overline{\text{Ocp}}(q)  = \frac{1}{4\pi} \sum_{V \in \Sigma} \left \langle \vec{n}(V),\vec{E_q}(V) \right \rangle \cdot \text{Area}^{*}(V),
\end{equation}
where \( V \) refers to a vertex. Alternatively, the discrete integration can be calculated as a facet-wise summation, which is more stable but consumes more memory:

\begin{equation}\label{eq:occupancy_facet}
    \overline{\text{Ocp}}(q) = \frac{1}{4\pi} \sum_{F \subset \Sigma} \left \langle \vec{n}(F),\vec{E_q}(c(F)) \right \rangle \cdot \text{Area}(F),
\end{equation}
where \( c(F) \) is the centroid of the facet \( F \).

\begin{figure}[!htp]
\centerline{\includegraphics[width=1\columnwidth]{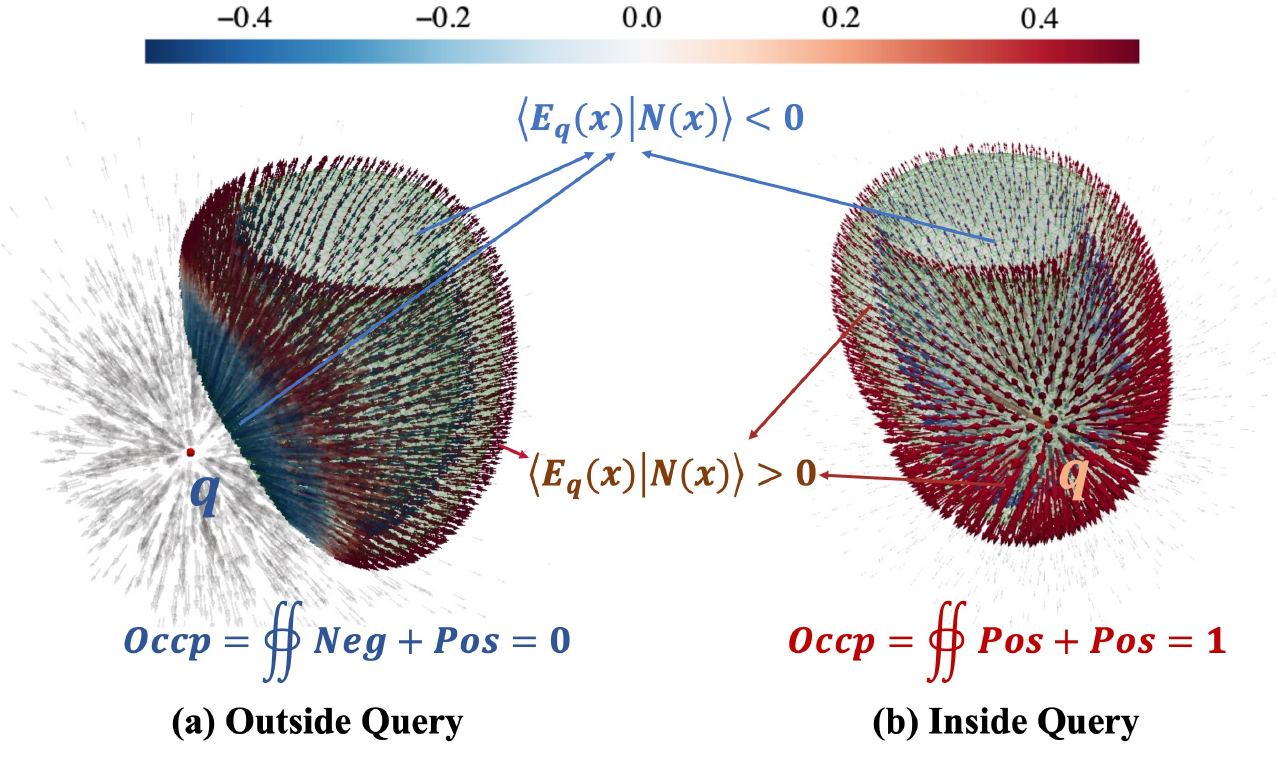}}
\captionsetup{justification=raggedright, margin=0.1cm}
\caption{\small The inverse quadratic fields derived from query points (a) outside and (b) inside a left ventricle mesh. The length of the arrows represents the field strength, and the color represents the inner product of the field and the outward normal vector of the mesh, blue signifies a negative value while red signifies a positive value.}
\label{fig:fields}
\end{figure}

To minimize the error of the discrete integration and maintain the continuous differentiability of the algorithm, we utilize the tanh function to approximate the binary value:

\begin{equation}\label{eq:occupancy_tanh}
    \text{Ocp}(q) = \tanh(\beta \cdot ( \overline{\text{Ocp}}(q) ) - \frac{1}{2}),
\end{equation}
where \( \beta \) is a hyper-parameter to control the smoothness of the approximation, usually set to \( 10^3 \) in our experiments. Fig. \ref{fig:fields} shows the inverse quadratic fields on the surface of a left ventricle mesh, derived from a query point inside the mesh and one outside, and the resulting occupancy value.

This formulation is differentiable as the operators used, inner products, summations, and continuous tanh activation functions are all differentiable. Furthermore, it is a global formulation as it considers the influence of a point source across all vertices or facets of the mesh. This contrasts with local formulations such as Chamfer loss and differentiable rasterization loss \cite{meng2023deepmesh}, which focus on localized mesh-to-voxel matching. Local formulations are prone to local minima, making convergence more challenging. 
The effects of the mesh-to-voxel alignment between the proposed differentiable voxelization algorithm with two implicit alternatives, ADSV \cite{joyce2022rapid} and Differentiable Rasterize \cite{meng2023deepmesh} is shown in Table \ref{table:MRcomp} in the next section. The center part demonstrates the comparison of the slicing accuracy. White masks the slicing of the original mesh. After the mesh is deformed, green shows the ground truth of the slicing of the updated mesh, and blue shows the slicing results from different methods.

The proposed algorithm is flexible and robust. Equation \eqref{eq:occupancy} provides an estimated occupancy even for non-watertight meshes. The DVS algorithm can be applied to point samples in the background space as well as inside the labeled cardiac space, and in combination, produce better results.


\subsection{Graph Harmonic Deformation (GHD)}

Our proposed GHD method is a model of the cardiac mesh. GHD describes the mesh deformation from a canonical or template mesh to the target mesh and is designed to preserve the mesh triangle quality and smoothness while keeping enough degrees of freedom to fully capture complex anatomic features.

GHD models the mesh connectivity as a graph, and mesh deformation as displacement vectors on the mesh nodes. Mesh deformation is modeled via Graph Fourier Transform (GFT), as a linear combination of the eigenfunctions of the Laplacian matrix of the cotangent-weighted mesh graph, each of which describes smooth and periodically fluctuating functions on the mesh surface. This effectively reconstructs the mesh surface via harmonic surface waves, and thus the name Graph Harmonic Deformation.

Mathematically, the graph Laplacian is defined as \( L = D - A \), where \( D \) is the degree matrix, a diagonal matrix with the degrees of each node on its diagonal, and \( A \) is the adjacency matrix of the graph. The eigenvectors of the Laplacian matrix provide a set of basis functions called the Graph Fourier Basis, denoted by \( U := [u_1, \dots, u_n]^T \), where \( L \cdot u_i = \lambda_i u_i \), \( 0 < i \leq N \). The eigenvectors corresponding to smaller eigenvalues of the graph Laplacian are the low-frequency Graph Fourier Basis functions, and vice versa. In this sense, the eigenvalues are the spectral energies, describing functions defined on the graph as fluctuant or smooth. After the basis is defined, GFT can be expressed by inner products of an arbitrary function with the bases. Mathematically, for a graph \( G \) and a real-valued function \( f : N_G \rightarrow \mathbb{R} \) on \( G \), we have the GFT of \( f \) as \( \phi = U^T \cdot f \), where \( U^T \) denotes the transpose of the Graph Fourier Basis \( U \), and \( \phi \) is called the Graph Fourier Coefficients. Conversely, the original function \( f \) can be naturally reconstructed by multiplying the Graph Fourier coefficients and the corresponding bases \( f = U \cdot \phi \). A natural smoothing filter on the graph can be formulated by preserving the first \( p \) low-frequency components of the Graph Fourier basis, i.e., \( f_p = \sum_{i < p} U_i \cdot \phi_i \).

We adopt the cotangent-weighted graph Laplacian, where weights are defined as:

\begin{equation}
 \omega_{ij} = \frac{\cot(\alpha_{ij}) + \cot(\beta_{ij})}{2},
\end{equation}

where \( \alpha_{ij} \) and \( \beta_{ij} \) represent the opposite angles of the edge \( e_{ij} \) joining vertices \( v_i \) and \( v_j \), in the two neighboring triangular faces respectively. The cotangent weights provide valuable information about the local geometry of the mesh, capturing the curvature and shape characteristics and the cotangent-weighted Laplacian is known as a typical approximation to the Beltrami-Laplacian of the base manifold \cite{rustamov2007laplace}. Our experiments show that the cotangent-weighted Laplacian facilitates the global deformation of the mesh and preserves the triangle quality and smoothness during the optimization without the need for additional regularization terms.

To balance various geometric information and avoid numerical issues caused by possible negative cotangent weights, i.e., \( \cot(\alpha_{ij}) + \cot(\beta_{ij}) < 0 \), we take the combined Laplacian from the above three kinds of weights to form a mixed Laplacian:

\begin{equation}
    L_{mix} = L_{cot} + \lambda_{norm}L_{norm} + \lambda_{unw}L_{unw},
\end{equation}
where \( L_{unw} \) is the unweighted Laplacian and \( L_{norm} \) is the inversely normalized Laplacian, the weight defined as \( w_{ij} = \frac{1}{\vert\vert x_j - x_i \vert\vert} \) viewed as the discretization of the directional derivative on the surface. Notice that the cotangent Laplacian occupies the main part of the mixed Laplacian, and the rest part of \( L_{norm}, L_{unw} \) can be regarded as the low-weighted regularization.

\begin{figure}[htbp!]
\centerline{\includegraphics[width=1\columnwidth]{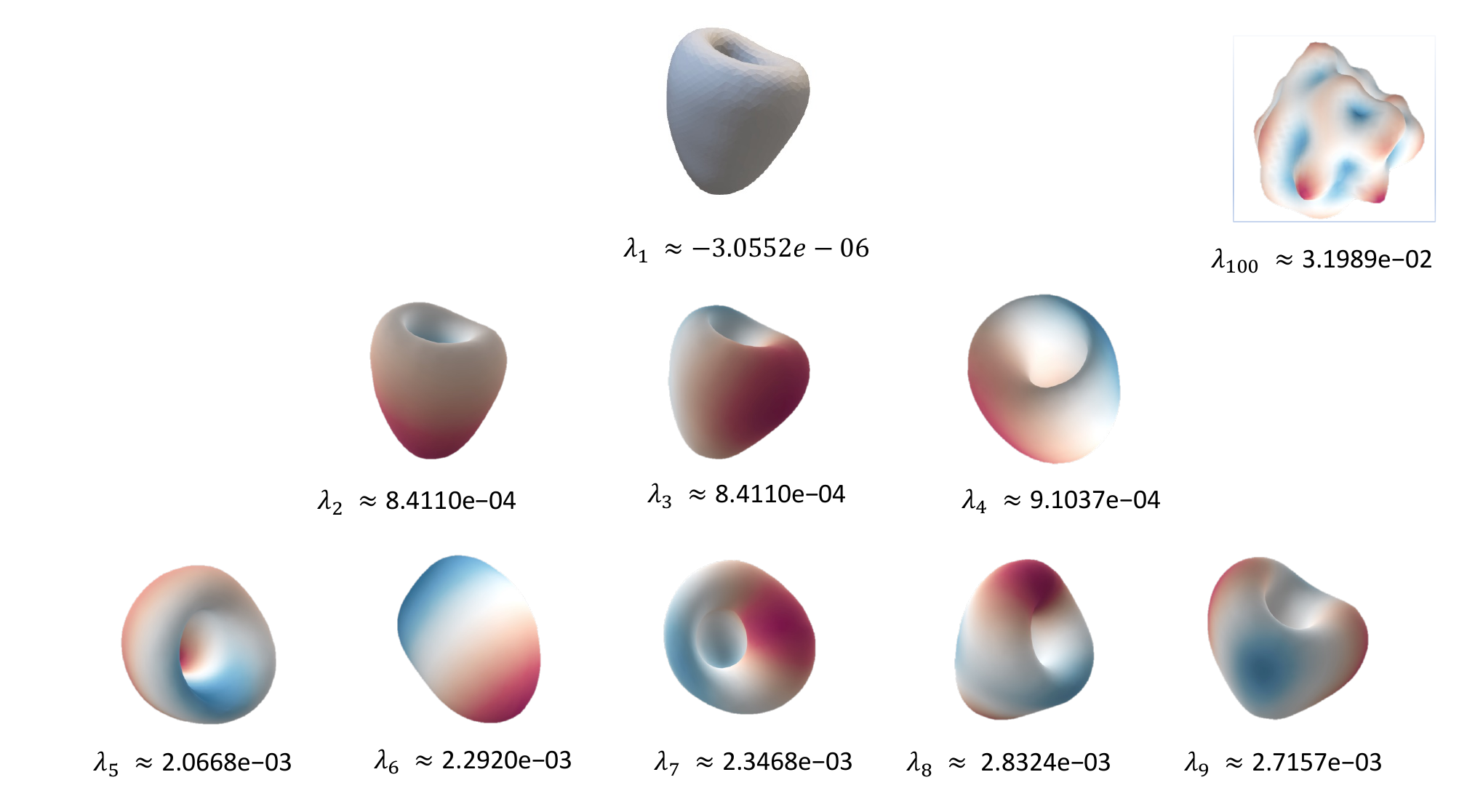}}
\captionsetup{justification=raggedright, margin=0.1cm}
\caption{\small The energy strips of the mixed Laplacian. The eigenvalues of the mixed Laplacian compose different energy strips rather than merely rank down to up. The higher energy strips represent the high-frequency components, which require more degrees of freedom to represent the mesh deformation. The Redshifts on each deformed LV mesh represent the positive (outward) displacements, and the blue for the negative (inward).}
    \label{fig:strip}
\end{figure}

Fig. \ref{fig:strip} demonstrates the energy strips of the various modes of the mixed Laplacian. The first mode represents the first wave number, with uniform energy across the surface; the second to fourth modes appear to be of the second wave number, where energy variations are monotonic across the surface in a half periodic surface function, while modes 5-9 appear to be of the third wave number, featuring a single periodic surface functions with a minima or maxima. As it appears that the number of modes at each wave number is in an arithmetic progression, with \( (2k-1) \) modes at the wave number of \( k \), the total number of modes employed should logically be \( \sum_{1}^{n} (2k-1) = k^2 \), where \( n \) is the number of different types of modes to be involved, so that no mode is missing from any particular wave number.

To achieve the desired shape, we engage in fitting GHD coefficients to minimize the loss between the deformed canonical mesh and the target. The optimization is effectively executed through Gradient Descent, represented mathematically as:

\begin{equation}
\Delta \phi = -\eta \cdot \frac{\partial}{\partial \phi} {\rm Loss}(\mathcal{M}_0 + U\phi, \hat{\mathcal{M}}),
\end{equation}

where \( \eta \) is the learning rate, \( \mathcal{M}_0 \) is the canonical mesh, \( \hat{\mathcal{M}} \) is the target mesh, and \( U\phi \) is the deformed mesh. When the mesh ground truth is available, the loss function can be defined as the Chamfer distance \cite{fan2017point} between the deformed mesh and the target mesh. When only the image-level supervision is possible, the loss function can be defined as the Dice loss \cite{milletari2016v} calculated from our differentiable slicing algorithm.

As the GHD naturally preserves the smoothness and quality of the mesh during the mesh morphing, there is no need for smoothness regularization constraints. Consequently, the GHD can focus on minimizing the target-oriented loss, and as such performs with higher efficiency to reach better convergence and accuracy than traditional mesh morphing approaches. Further, the natural smoothness enables bridging between sparse image sampling, allowing the GHD to remain robust even when only very few image planes or image voxels are available.

\subsection{Differentiable Physiologic Constraints}

Besides voxelization and slicing, we propose a few differentiable mesh operations for auxiliary supervision, which can be implemented in 3D and 4D cardiac reconstruction tasks. Several others are possible.  


\subsubsection{Differentiable Thickness}

During the fitting of sparsely sampled image data, controlling the thickness of the cardiac wall can be difficult, and crossover of the inner and outer surfaces can happen due to insufficient image supervision. This is especially so at the apex of the heart, where the inner and outer surfaces are topographically very far away. To address this, we introduce a differentiable thickness function that can be used to regularize thicknesses, such as having a minimum or adopting a value similar to elsewhere in the cardiac chamber.

\begin{equation}\label{eq:thickness}
\text{Thickness}(\vec{q}) = \min_{\vec{p} \in \mathcal{M}} \left( \left\Vert \vec{q} - \vec{p} \right\Vert + \lambda \left\Vert \vec{N_p} + \vec{N_q} \right\Vert \right),
\end{equation}
where \( \vec{q} \) is the query point on the mesh, \( \vec{p} \) is the closest point on a face on the opposite side of the thickness, found via the differentiable point\_face\_distance\_forward PyTorch function, \( \vec{N_p} \) and \( \vec{N_q} \) are the normal vectors of the faces containing \( \vec{p} \) and \( \vec{q} \) respectively, and \( \lambda \) is a hyper-parameter to balance the distance and the normal consistency. The formulation assumes that the two points defining the local wall thickness are at minimum distance apart and have the best-aligned face normals. The thickness function can be optimized by the gradient descent algorithm, and it can be used to formulate a traditional mean square error loss term to supervise mesh reconstruction.

\subsubsection{Volume \& Weak Incompressibility}

We further introduce a differentiable volume function for supervising mesh scaling and deformation, as well as a differentiable change of volume function to weakly enforce myocardial incompressibility physics. The differentiable function is given as:

\begin{equation}\label{eq:volume}
V = \iiint_{\Omega} \nabla \cdot \vec{x} \, \mathrm{d}v = \iint_{\partial \Omega} \langle \vec{n}, \vec{x} \rangle \, \mathrm{d}s,
\end{equation}
where \( \vec{x} \) is the position vector and \( \vec{n} \) is the normal vector of the mesh \( M \). The volume can be converted to the surface integral by the divergence and Gauss theorem. The discretization of Equation \eqref{eq:volume} is given as:

\begin{equation}\label{eq:volume_discrete}
V = \frac{1}{3} \sum_{F \subset \Sigma} \langle \vec{n}(F), \vec{x}(c(F)) \rangle \cdot \text{Area}(F),
\end{equation}
where \( c(F) \) is the centroid of the face \( F \). The differentiable volume can be used to supervise the mesh scaling and the deformation of the myocardium.

To enforce incompressibility, we use a function describing the change of volume within the mesh, which can be enforced to be zero in a loss term, expanded via the chain rule:

\begin{equation}\label{eq:incompressible}
\begin{aligned}
\frac{\mathrm{d} V}{\mathrm{d} t} &= \frac{\mathrm{d}}{\mathrm{d} t} \iint_{\Sigma} \langle \vec{n}, \vec{x} \rangle \, \mathrm{d}s = \iint_{\Sigma} \frac{\mathrm{d}}{\mathrm{d} t} \langle \vec{n}, \vec{x} \rangle \, \mathrm{d}s \\
&= \iint_{\Sigma} \left\langle \frac{\mathrm{d}}{\mathrm{d} t} \vec{n}, \vec{x} \right\rangle \, \mathrm{d}s + \iint_{\Sigma} \left\langle \vec{n}, \frac{\mathrm{d}}{\mathrm{d} t} \vec{x} \right\rangle \, \mathrm{d}s \\
&\simeq \iint_{\Sigma} \left\langle \Delta_{t}^{t+1} \vec{n}, \vec{x} \right\rangle \, \mathrm{d}s + \iint_{\Sigma} \left\langle \vec{n}, D_{t}^{t+1} \right\rangle \, \mathrm{d}s = 0,
\end{aligned}
\end{equation}
where \( \Delta_{t}^{t+1} \vec{n} \) is the difference of the normal vector between time \( t \) and \( t+1 \), and \( D_{t}^{t+1} \) is the mesh displacement field at time \( t \) to \( t+1 \).

The weak incompressible constraint will be applied to reconstruct 4D cardiac motion.

%% file: results/exp.tex
\section{Proposed GHD-DVS Framework }

In this section, we present the GHD-DVS pipeline, its loss functions, the datasets used for experiments and the performance measures used to evaluate the results.

\subsection{Overall Pipeline}

\begin{figure*}[t!]
\begin{center}
        \includegraphics[width = \textwidth]{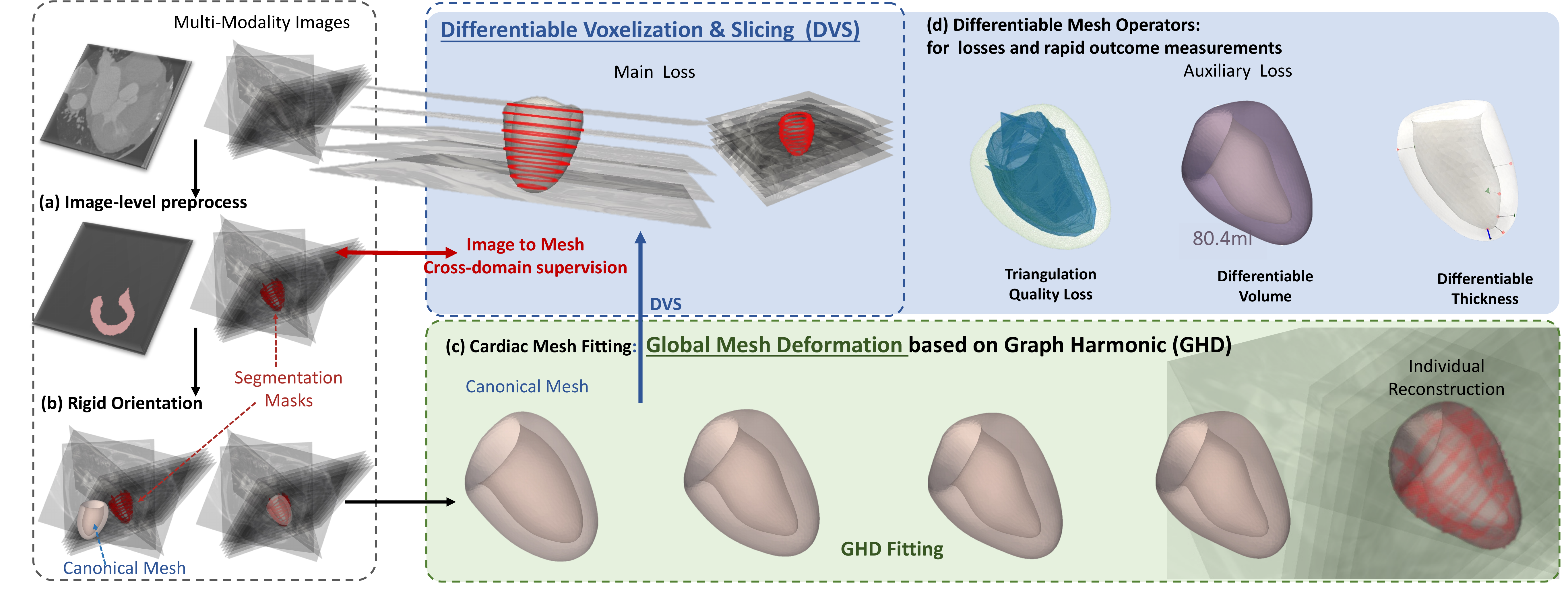}
        \end{center}
    \captionsetup{justification=raggedright, margin=0.1cm}
    \caption{\small The pipeline of the 3D mesh reconstruction of the left ventricle from MRI images. The pipeline starts from the segmentation of medical images, followed by a rigid orientation to align the canonical shape to the target roughly. The GHD optimization supervised by the differentiable slicing on the images yields the reconstructed mesh. Auxiliary supervisions like thickness and volume are alternatively applied to improve the mesh quality for better biomedical plausibility.}
    \label{fig:pipeline}
\end{figure*}

The overall pipeline for the differentiable mesh reconstruction combining DVS and GHD is shown in Fig. \ref{fig:pipeline}. The process starts with image-level pre-processing, where the raw images are de-blurred, resampled, and intensity normalized as is typically done \cite{mcauliffe2001medical}. The heart region is then segmented from the images, and this can be performed manually or via a deep learning algorithm, such as U-Net in \cite{siddique2021u}. The labeled pointclouds from the right ventricle, left ventricle, left ventricle cavity and anocelia are then extracted from the segmented images. It is worth noting that the labeled point clouds are sparse and irregularly distributed when the raw images stack is sparse. In our experiments, we directly extracted the labeled point clouds from the manual annotation provided in the dataset to avoid potential errors from the segmentation model.

Before the mesh reconstruction, the canonical shape is aligned to the target image roughly.
The rigid orientation from the canonical mesh to the target pointclouds is obtained by optimizing the quaternion representation of the rotation matrix \cite{altmann2005rotations} and the translation vector among the random sampling points from the canonical shape and the target image stack. The right ventricle alignment is considered in the rigid orientation, which facilitates breaking the symmetry of the left ventricle to avoid wrong-paired fitting during the GHD optimization (shape fitting converging well while indices are misaligned).

The canonical shape is a manually optimized LV mesh with 4000 vertices approximating the mean shape of the training dataset. Our experiments show that our mesh morphing method can be sensitive to the canonical shape chosen, however, due to the approximately axisymmetric shape of the left ventricle, it is not difficult to find a canonical shape that works well. However, for more complex geometries such as cranial aneurysms attached to surrounding blood vessels, a good canonical shape close to the average of anticipated geometries is required.

The entire pipeline includes the rough rigid orientation and the GHD optimization supervised by the differentiable slicing. The optimization is executed by the Adam optimizer \cite{kingma2014adam}. The rigid orientation from the canonical mesh to the target is obtained by optimizing the quaternion representation of the rotation matrix and the translation vector among the random sampling points from the canonical shape and the target image stack. Notice that the right ventricle alignment is considered in the rigid orientation, which facilitates breaking the symmetry of the left ventricle to avoid wrong-paired fitting during the GHD optimization. See Fig. \ref{fig:pipeline} for the whole pipeline of our 3D mesh reconstruction.

\subsection{Loss Functions}

We adopt the Dice loss (\cite{milletari2016v}) to supervise the match between the reconstructed mesh and ground truth image segmentations:

\begin{equation}
    \text{Dice}(\mathcal{M}_t, \hat{Ms}) =  \frac{2\sum_{P_i \in \text{Samples}} \overline{Ocp}(P_i) \cdot \hat{Ms}(P_i)
    }{\sum \overline{Ocp}(P_i) + \sum \hat{Ms}(P_i)},
\end{equation}
where \( \overline{Ocp}(P_i) \) is the occupancy of samples \( P_i \) toward the current mesh, and \( \hat{Ms}(P_i) \) is the ground truth mask from the slicing of the images. This loss term enforces the occupancy of all segmented voxels in the image planes to be within the mesh, and the non-occupancy of all non-segmented voxels to be outside the mesh.

Since the GHD is naturally smooth, we do not need to regularize smoothness. The only regular term we use is the thickness constraint to avoid zero-thickness and mesh collapse at the apex during reconstructions of sparse image data. We constrained apical thickness to be greater than 4 mm, as informed by thickness reports of left ventricle thickness in the adult population \cite{walpot2019left}, using the thickness loss:

\begin{equation}
    \text{Loss}_{th} = \sum_{P_i \in \text{Samples}} \text{SiLU}(\text{Thickness}(P_i) - 4 \, \text{mm}),
\end{equation}
where \( \text{SiLU}(x) = \frac{x}{1 + e^{-x}} \) \cite{hendrycks2016gaussian} is the scaled linear unit function. The final loss function is the combination of the Dice loss and the thickness loss:

\begin{equation}
    \text{Loss} = \text{Dice} + \lambda \cdot \text{Loss}_{th},
\end{equation}
where \( \lambda \) is the weight of the thickness loss. The weight of the thickness loss is set to 0.01 in our experiments.

\subsection{Datasets}

In our study, we employed and analyzed cardiac ventricular volumetric structural imaging data derived from two computed tomography (CT) datasets, MMWHS \cite{zhuang2019evaluation} and CCT48 \cite{suinesiaputra2017statistical}, and three magnetic resonance (MR) datasets, ACDC \cite{bernard2018deep}, UK Biobank (UKBB) \cite{petersen2016uk}, and the MITEA dataset \cite{zhao2023mitea}. The CT datasets, MMWHS and CCT48, are characterized by their high-resolution images, with an in-plane resolution of 0.78 mm and a slice thickness of 1.6 mm, making them dense imaging data sources due to the finely detailed spatial resolution they provide. The MR datasets, ACDC and UK Biobank, exhibit a sparser imaging structure with larger gaps between imaging planes. Specifically, ACDC has an in-plane resolution of 1.37 to 1.68 mm with slice thicknesses varying between 5 and 10 mm. The UK Biobank provides MR imaging data with a resolution of 1.7 mm, with slice thicknesses of 6.0 mm in short-axis (SAX) views and 8.0 mm in long-axis (LAX) views. The MITEA dataset is a set 3D echocardiography data that is annotated via multi-modality reconstruction with matching MRI images.

MMWHS comprises 20 patients, CCT48 comprises 48 patients, and we utilized data for 100 patients from ACDC, data for 16 patients from UK Biobank, and data for 10 patients from MITEA.

\subsection{Performance Measures}

We evaluated the performance of our method using several metrics. The Dice coefficient was used to measure the overlap between the reconstructed mesh and the ground truth segmentations. The Chamfer Distance (CD) and Hausdorff Distance (HD) were used to evaluate the accuracy of the reconstructed mesh's surface compared to the ground truth surface.

The Chamfer Distance (CD) between the reconstructed mesh and the ground truth surface is defined as:

\begin{equation}
\label{eq:CD}
    \text{CD}(\mathcal{M}, \hat{\mathcal{M}}) = \frac{1}{|\mathcal{M}|} \sum_{x \in \mathcal{M}} \min_{y \in \hat{\mathcal{M}}} \| x - y \|^2 + \frac{1}{|\hat{\mathcal{M}}|} \sum_{y \in \hat{\mathcal{M}}} \min_{x \in \mathcal{M}} \| y - x \|^2,
\end{equation}
where \( \mathcal{M} \) and \( \hat{\mathcal{M}} \) are the reconstructed and ground truth meshes, respectively.

The Hausdorff Distance (HD) is given by:

\begin{equation}
\label{eq:HD}
    \text{HD}(\mathcal{M}, \hat{\mathcal{M}}) = \max \left\{ \sup_{x \in \mathcal{M}} \inf_{y \in \hat{\mathcal{M}}} \| x - y \|, \sup_{y \in \hat{\mathcal{M}}} \inf_{x \in \mathcal{M}} \| y - x \| \right\}.
\end{equation}

These performance measures provide a comprehensive evaluation of the reconstructed mesh's accuracy and fidelity to the ground truth.

\section{Results}
\subsection{Convergence and Robustness of GHD}
\label{subsection4_1}

\begin{figure}[h!]
\centerline{\includegraphics[width=1\columnwidth]{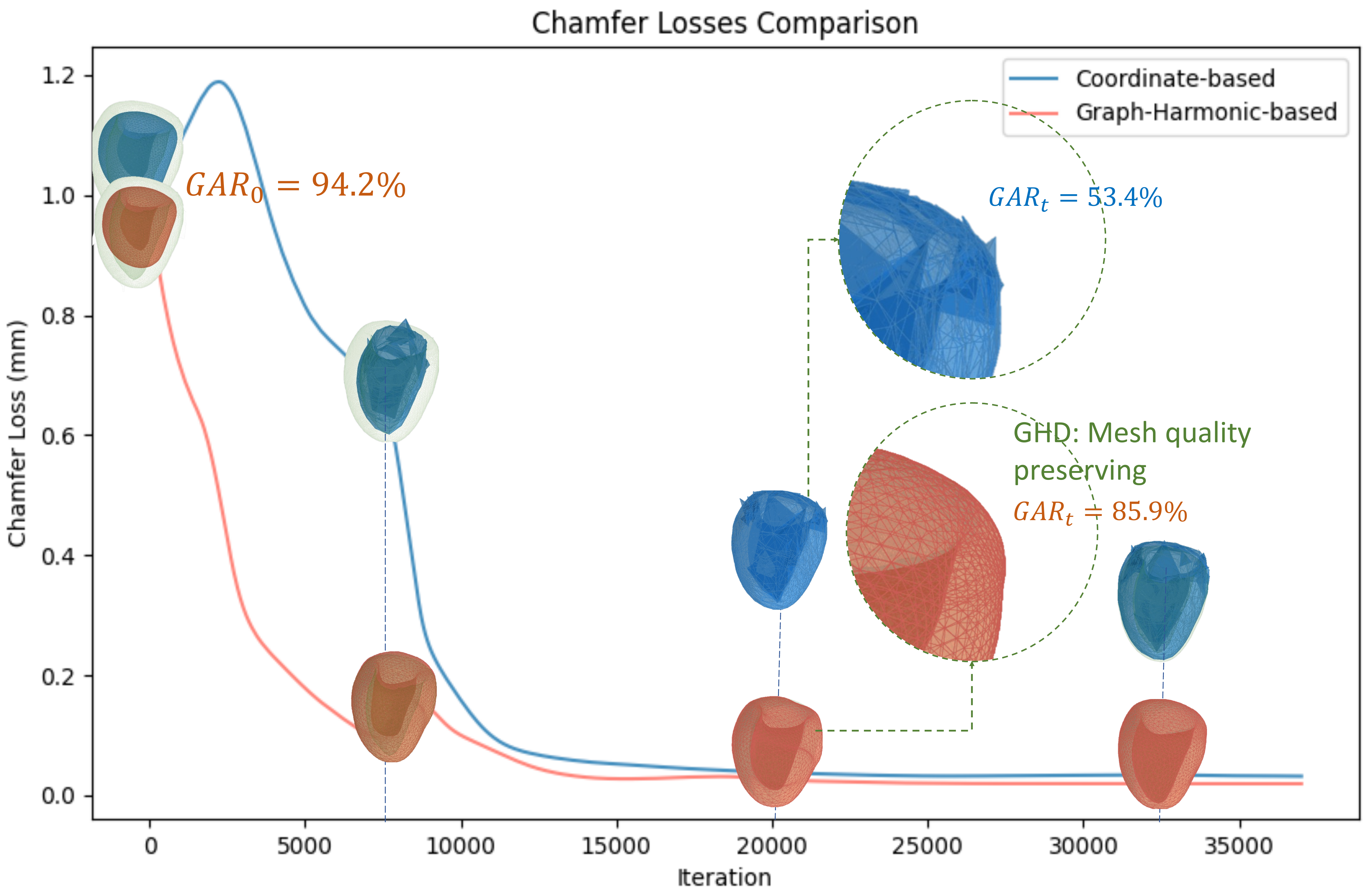}}
\captionsetup{justification=raggedright, margin=0.1cm}
\caption{\small The convergence of GHD compared to the traditional vertex-wise mesh morphing method, demonstrating that GHD converges faster and more robustly. The zoomed-in view of the mesh shows that the GHD method preserves the triangle quality and smoothness during the optimization, while the vertex-wise deformation method has irregular triangles and sharp edges during optimization.}
\label{fig:convergence}
\end{figure}

We first evaluate the advantages of using the GHD to reconstruct the left ventricle myocardial mesh via mesh morphing, supervised by a smooth ground truth mesh, using Chamfer distance as the loss function. We compare GHD to a traditional vertex-wise formulation (where mesh deformation is modeled directly as vertex displacements). The results in Fig. \ref{fig:convergence} (a) show that with the same optimizer and learning rate, the GHD method converges better with a lower Chamfer loss, demonstrating efficiency and robustness. Furthermore, visual observations demonstrate that the GHD preserves the triangle quality and smoothness during the optimization better, as the vertex-wise method results in a mesh with irregular triangles and sharp edges. With stronger smoothness regularization, such irregularities on the vertex-wise approach can reduce, but the convergence becomes poorer and ends with a higher Chamfer loss. With natural smoothness, the GHD avoids such a trade-off.

Preserving mesh qualities is the essential advantage of GHD over vertex-wise displacement. We use the radio of triangles only containing good angles (between 30 and 120 degrees) as a measure of mesh quality to avoid extremely acute or obtuse triangles, which causes numerical instability in downstream applications, like finite element simulations. The good angle ratio (GAR) is defined as:
\begin{equation}
    \text{GAR} = \frac{| \{T\in F| \forall \theta \in T, \frac{1}{6}\pi \leq \theta \leq \frac{2}{3}\pi\} |}{| F |},
\end{equation}
where \(F\) is the set of all triangles in the mesh, and \(\theta\) is the angle of any triangle \(T\). The results in Fig. \ref{fig:convergence} show that GHD preserves mesh quality during optimization, while coordinate-based morphing decreases the GAR significantly, from 0.942 to 0.534, indicating that the mesh quality is degraded during optimization.

\begin{figure*}[t!]
    \centering
    \captionsetup{justification=raggedright, margin=0.1cm}
    \includegraphics[width=1\linewidth]{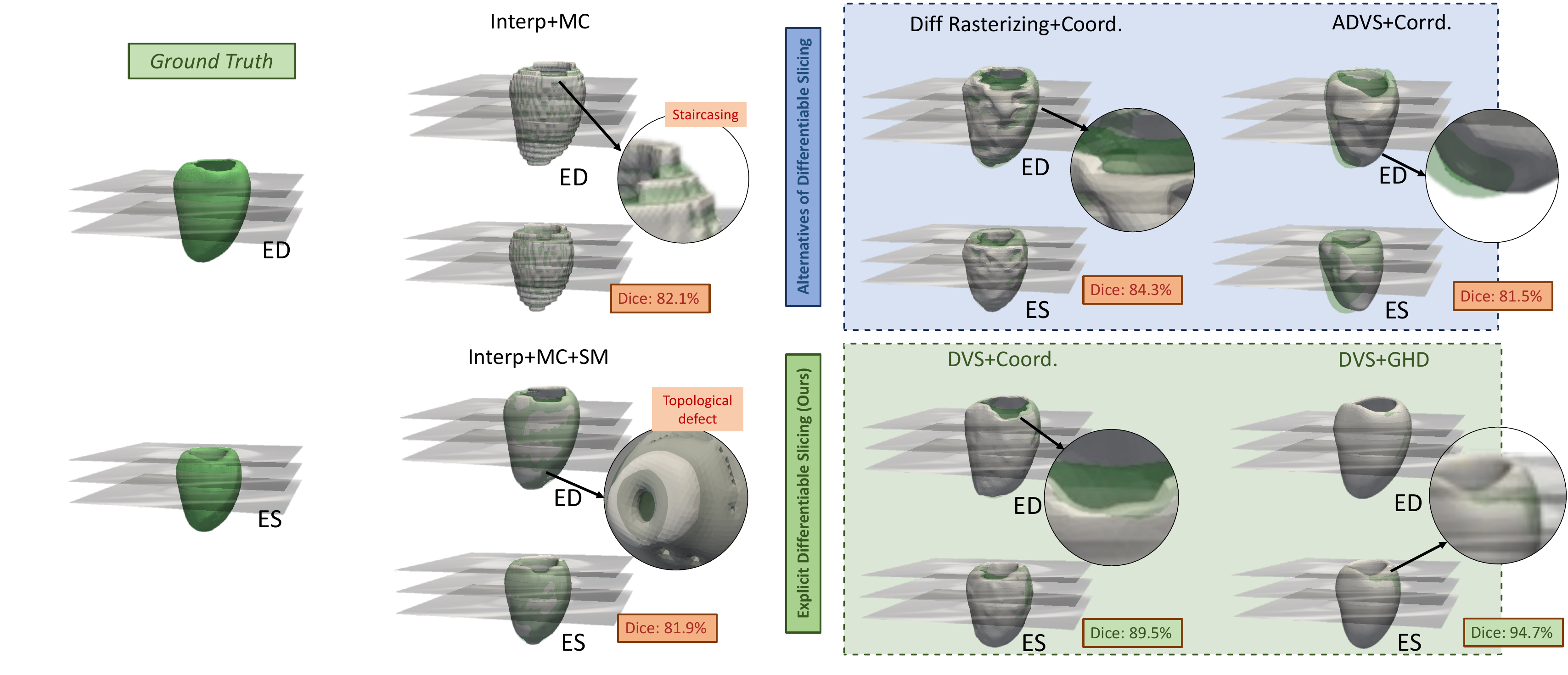}
    \caption{\small The comparison of LV reconstructions from sparse MRI images using different methods. The green mesh is the ground truth from the manual annotation on high-resolution images, and the white meshes are the reconstructed meshes from the sparse images using different methods. The results show the reconstructions for the phase of end-diastolic (ED) and end-systolic (ES) respectively.}
    \label{fig:out}
\end{figure*}

\subsection{Performance of GHD-DVS on Dense Image Data (CT)}

We compare GHD to coordinate-based mesh morphing algorithms and statistical shape models (SSM) on their effectiveness and robustness as a shape representation by testing them on 3D mesh reconstruction on dense CT images. SSMs evaluated include models utilizing PCA of node coordinates as basis modes, and a model utilizing a VAE of node coordinates for reconstruction. for PCA SSM, we investigate two models, one trained by us using limited number of CT images and another from \cite{bai2015bi}, which is trained with 1000 MR images. Unlike mesh morphing approaches (GHD and coordinate-based mesh morphing), SSM requires pre-training with a dataset. The self-trained PCA SSM and the VAE SSM are trained with 20 samples from the MMWHS dataset, while the Bai et al. PCA SSM is adopted in the trained state. 20 modes are used from each PCA during reconstruction. 10 cases from MMWHS and the 48 samples from the CCT48 dataset are used for testing all methods.

The evaluation task is to fit the various models to the ground truth mesh by minimizing the learnable parameter, \( \theta \) (mode or latent vector weights for PCA or VAE, GHD weights for GHD, and vertex displacements for coordinates-based mesh morphing), via the gradient descent optimizer, as follows:
\begin{equation}
    \label{eq:gradient descent}
    \Delta \theta = - \eta \nabla_{\theta} \text{Loss}(\mathcal{M}, \hat{\mathcal{M}}),
\end{equation}
where \( \eta \) is the learning rate, and \( \text{Loss} \) is either CD or DVS, $M$, $\hat{M}$ are the optimizing mesh and the ground truth mesh, respectively.

Here, a ground truth mesh is required as CD can only be evaluated between two meshes. We utilized the unsmoothed marching cubes mesh as the ground truth mesh as it can represent the labelled image. Its shortcomings of having staircasing artefacts or imperfect mesh quality does not prevent the evaluation of the geometric accuracy of various shape reconstruction algorithms. 
We evaluate the performances using the Dice score (proportion of labelled image voxels within the mesh), and CD and HD between the reconstructed and ground truth meshes, calculated via Eq.\eqref{eq:CD} and Eq.\eqref{eq:HD}.

 

\begin{table}[htbp!]
\caption{Performance of various algorithms as shape representations, after reconstruction on CT images. MC+SM - marching cubes with smoothing; Coord - Coordinate-based mesh morphing.}
\label{table:CTcomp}
\renewcommand\arraystretch{1.3}
\centering
\setlength{\tabcolsep}{2pt}
\resizebox{1.0\columnwidth}{!}{
\scriptsize
\begin{tabular}{l|c|c|c}
\toprule[0.1mm]
\hline
Methods                & Dice (\%)  & CD (mm) & {HD (mm)}  \\ \hline
MC+SM (CD)  & 98.5 (0.6)   &   0.05(0.03)    &  0.92(1.7) \\ \hline
Coord (CD)  & 97.2 (1.2)   & \textbf{$<$0.01}       & 1.8(0.8) \\ \hline
PCA (self-trained) (CD)  & 81.6 (2.1)   & 0.52 (0.3)        & 3.6(0.4) \\ \hline
PCA (\cite{bai2018automated}) & 79.1 (3.5)   & 0.47 (0.2)        & 2.6(0.3) \\ \hline
VAE (Training) (CD) & \textbf{99.2 (0.4)} & \textbf{$<$0.01} & 0.29(0.1) \\ \hline
VAE (Testing) (CD) & 76.0 (3.7) & 3.17 (0.43) & 7.5 (2.6) \\ \hline
GHD (CD) & 93.1(1.5) & \textbf{0.02 (0.01)} & \textbf{0.16 (0.07)} \\ \hline
Coord (DVS)  & \textbf{99.6 (0.1)}   & 0.25(0.09)      & 3.2(1.1) \\ \hline
GHD (DVS) & \textbf{98.6 (0.8)} & 0.13 (0.04) & 2.1(0.2) \\ \hline
\end{tabular}}
\end{table}

Results are shown in Table \ref{table:CTcomp}. 
Mesh morphing methods (coordinates-based and GHD) generally outperformed SSMs (PCA and VAE) in all measures. Modal constraint in PCA SSM may be reducing the goodness of the fit between reconstructed mesh and image labels. The two PCA SSM models performed similarly, suggesting that the low number of training cases is not limiting the performance of our self-trained PCA SSM. The VAE SSM appeared overtrained, with good performance during training but poor performance during testing, which may be associated with the low sample size available for training. We acknowledge that it may improve with a larger training dataset but do not have a larger training dataset for further investigations. However, this result points to the advantage of GHD, which does not require any pretraining or a training dataset for reconstruction.

Comparing GHD to coordinates-based mesh morphing approaches, GHD has lower CD and HD, suggesting that it has a smoother and better quality mesh, corroborating results in subsection \ref{subsection4_1}, but GHD has lower Dice, which is consequent to GHD being a lower dimensional representation of shape. However, with DVS, dice performance of GHD is insignificantly lower than that of coordinates-based mesh morphing. Further, comparing GHD to marching cubes with smoothing, Dice and CD performance are very similar, although GHD has slightly higher HD.

The results thus suggest that GHD is a good method for mesh reconstruction on dense images, as it concurrently offers good accuracy and good mesh qualities. GHD enables a reduced dimensional representation of shape without sacrificing the ability to accurately model an unseen shape, and it does not require pre-training with a large dataset. Unlike marching cubes, GHD provides reconstructions where the number of vertices are fixed, and may be easier to use for downstream processes such as motion quantifications. 

Comparing CD to DVS for GHD and coordinates-based mesh morphing approaches, we observe that DVS gives better Dice while CD gives better CD and HD. This is naturally so as the CD loss attempts to minimize distances between mesh, while DVS attempts to optimize fit with image labelled voxels.

\subsection{Performance of GHD-DVS on Sparse Image Data (MRI)}

The 3D mesh reconstruction for MRI short-axis stack images is more challenging as image data is sparser, due to large spacing between image planes. We tested various mesh reconstruction algorithms on the ACDC and UKBB datasets, where the MRI images is scanned at 5 to 20 slices. The MRI datasets regularly contains bad cases with misalignment between the image slices. Therefore, during this experiment, we manually selected 15 cases from each dataset with good alignment. 
All cases are down-sampled to 5 short-axis slices for the sparse MRI reconstruction experiment. Reconstructions are performed only from the short-axis stack, but evaluation of Dice (percentage occupancy of labelled voxels within mesh) is performed on both short axis and long axis (including 2 and 4 chamber view) image slices,

We compared GHD+DVS to several algorithms: (1) marching cubes applied to the interpolated MRI image (interpolation conducted to convert to isotropic resolution) with or without smoothing, (2) several mesh morphing approaches that utilize the vertex-wise displacement morphing model. These mesh morphing approaches include those that use ADVS \cite{joyce2022rapid}, differentiable Rasterization \cite{meng2023deepmesh}, and our DVS as the image guidance of mesh reconstruction. The comparison is shown in Table \ref{table:MRcomp}.

Results show that mesh marching cubes without smoothing achieves only moderately good Dice, CD and HD performance. However, the mesh quality is poor with staircasing artefacts, and the global topology is sometimes not controllable, with holes and other topological defects appearing, an example is shown in the second column of Fig. \ref{fig:out}. While staircasing artefacts can be removed by smoothing, topology defects are hard to fix. Combining smoothing post-processing with marching cubes does not significantly improve performance measures, but it improves the visual quality of the mesh.

On the average, coordinates-based mesh morphing approaches do not do substantially better than the unsmoothed marching cubes approach. However, the DVS-based mesh morphing performs better than marching cube in almost all measures. Comparing the 3 techniques for image guidance, Differentiable Rasterization performs better than ADVS, achieving both higher Dice and lower CD and HD, but it does not perform as well as DVS. The results thus suggest that DVS is a superior image guidance technique during mesh reconstruction. 

Finally, comparing coordinates-based mesh morphing guided by DVS to GHD guided by DVS, we observe further and substantial improvement in Dice, mild improvement in HD, and a slightly poorer but very similar CD. This shows that GHD is a better technique than direct vertex morphing for mesh reconstruction on sparse images. Further, GHD+DVS achieved the best results in all the methods tested, showing that it can outperform the state of the art.

\begin{table*}[t!]
\caption{Performance of various algorithms in 3D mesh reconstruction from Sparse (5-slice short axis) MRI images. Interp - image interpolation pre-processing; SM - smoothing postprocessing; MC - marching cubes; Coord - coordinates-based mesh morphing; ADVS - \cite{joyce2022rapid}, Diff Rast - \cite{meng2023deepmesh}}
\label{table:MRcomp}
\resizebox{\textwidth}{!}{
\setlength{\tabcolsep}{1.2pt}
\centering
\renewcommand\arraystretch{1.3}
\begin{tabular}{c|ccc|cccc|ccccc}
\toprule[0.1mm]
\hline
\multirow{3}{*}{Methods} & \multicolumn{3}{c|}{Properties}                                             & \multicolumn{4}{c|}{ACDC}                                                                                        & \multicolumn{5}{c}{UKBB}                                                                                            \\ \cline{2-13} 
                         & \multicolumn{1}{c|}{\scriptsize Uniform} & \multicolumn{1}{c|}{\scriptsize Mesh}    & \scriptsize Smoothing & \multicolumn{2}{c|}{Dice (\%)}             & \multicolumn{1}{c|}{\multirow{2}{*}{CD (mm)}} & \multirow{2}{*}{HD (mm)} & \multicolumn{3}{c|}{Dice (\%)}                          & \multicolumn{1}{c|}{\multirow{2}{*}{CD (mm)}} & \multirow{2}{*}{HD (mm)} \\ \cline{5-6} \cline{9-11}
                         & \multicolumn{1}{c|}{\scriptsize Mesh}    & \multicolumn{1}{c|}{\scriptsize Quality} & \scriptsize Required      & 3D         & \multicolumn{1}{c|}{SAX} & \multicolumn{1}{c|}{}                         &                          & 3D         & SAX        & \multicolumn{1}{c|}{LAX} & \multicolumn{1}{c|}{}                    &                     \\ \hline
Interp+MC                & F                            & B                            & F             & 84.1(1.6) & 80.7(5.2)               & 0.82(0.13)                                   & 1.76(1.2)                & 82.1(0.9) & 85.4 (2.4) & 78.8(2.5)               & 1.1(0.9)                                 & 3.51(1.7)           \\ \hline
Interp+MC+SM              & F                            & G                            & T             & 83.7(1.5) & 81.6(3.4)               & 0.83(0.20)                                   & 1.13(0.8)                & 81.9(0.9) & 86.1 (2.8) & 79.3(2.3)               & 1.5(1.2)                                 & 2.42(1.8)           \\ \hline
Coord.+ADVS                     & T                            & B                            & T             & 79.7(2.4)          & 80.7(1.8)                        & 1.1(0.20)                                             & 3.8(1.2)                        & 81.5(4.6)          & 84.4(3.9)          &    79.8(2.8)                & 0.5(0.16)                                        & 4.82(1.6)                   \\ \hline
Coord.+Diff Rast.          & T                            & B                            & T             & 81.3(1.4)         & 82.6(2.1)                        & 0.75(0.29)                                        & 2.14(0.4)                 & 84.3(1.8)          & 85.3(3.6)          &    87.6(2.9)                & 0.14(0.04)                                        & 3.23(0.1)                                    \\ \hline
Coord.+DVS                      & T                            & B                            & T             & 83.5(0.6)         & 81.4(4.9)                    & \textbf{0.09(0.02)}                                             & 1.16(0.5)                        & 89.5(0.6)       & 87.4(2.3)       & 93.4(2.3)                    & \textbf{0.04(0.01)}                                     & 0.73(0.4)           \\ \hline
GHD+DVS                  & T                            & \textbf{G}                            & \textbf{F}             & \textbf{90.8(0.6)}         & \textbf{94.2(1.2)}                       & 0.11(0.01)                                            & \textbf{0.92(0.2)}                       & \textbf{94.7(0.2)}         & \textbf{90.8(2.0)}        & \textbf{97.8(0.7)}                        & 0.06(0.01)                                        & \textbf{0.54(0.1)}                   \\ \hline
\end{tabular}}
\end{table*}

From the visual results in Fig. \ref{fig:out} and Table \ref{table:MRcomp}, we can see that GHD+DVS can reconstruct the left ventricle mesh from sparse MRI images with high accuracy and fidelity to the ground truth. The reconstructed meshes are visually similar to the ground truth meshes, with good alignment and smoothness. The results demonstrate the effectiveness of GHD+DVS for 3D mesh reconstruction from sparse MRI images.

\subsection{Clinical Analysis}
To validate the clinical applicability of GHD+DVS, we employed it to compute left ventricular (LV) volume, ejection fraction (EF), and myocardial strains. The end-diastolic volume (EDV) was determined after fitting both the marching cubes (MC) method and GHD+DVS to the same ground truth segmentations. We utilized 50 segmentations from the MITEA 3DE dataset (\cite{zhao2023mitea)} which comprises annotated 3D echocardiography scans with segmentations of the LV myocardium and cavity derived from paired cardiac MRI scans. To assess each method's ability to reconstruct the LV mesh with reduced ground truth labels, the segmentations were progressively reduced.

\begin{figure*}[t!]
    \centering
    \captionsetup{justification=raggedright, margin=0.1cm}
    \includegraphics[width=1\linewidth]{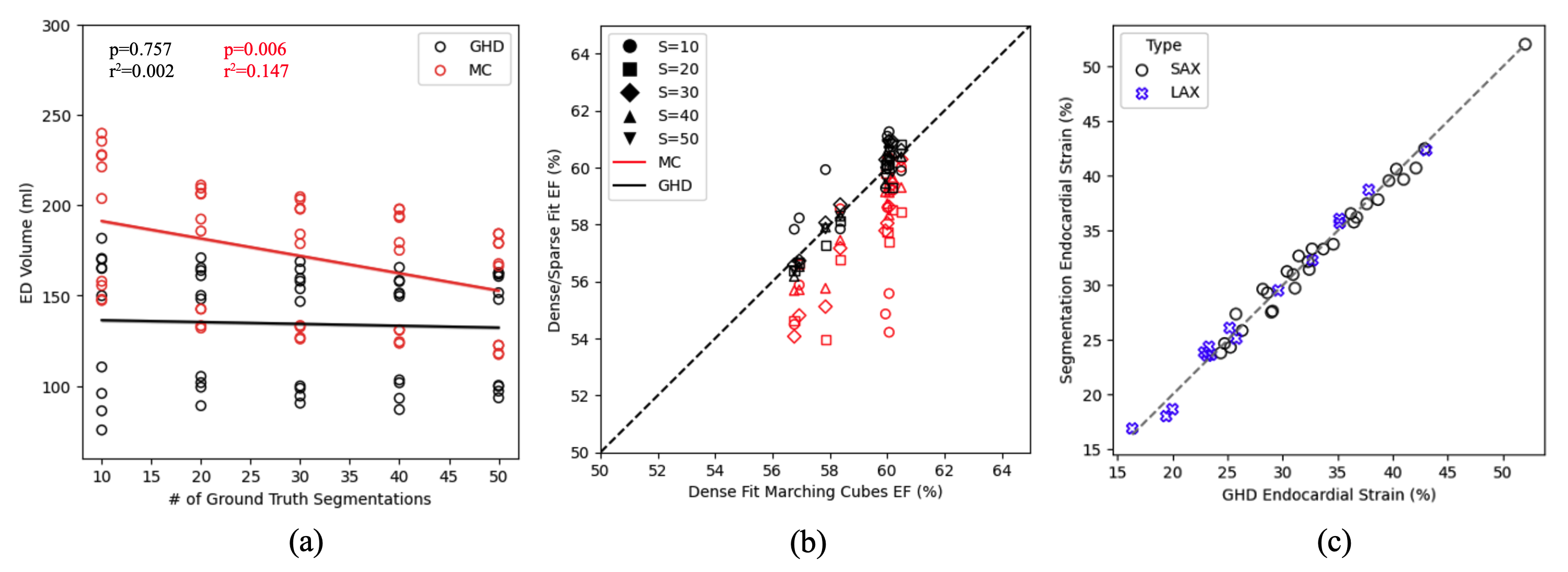}
    \caption{\small Clinical analysis results. (a) End diastolic (ED) volume measured from the resultant marching cubes (MC) and GHD fit to dense data (segmentations=50) and reduced ground truth labels (segmentations = 40 and below). (b) EF quantification comparison between dense fit MC (Gold Standard) and both MC and GHD fit to sparse segmentations. For volume and EF, 10 cases from the MITEA 3DE dataset was used. (c) Strain measurements comparison between GHD+DVS and manual calculations at the endocardial border measured across 15 UK Biobank cases.}
    \label{fig:clinical_analysis}
\end{figure*}

For EDV and EF analyses (refer to Fig. \ref{fig:clinical_analysis}(a \& b)), we examined 10 cases, varying the number of slices from 10 to 50, ranging from sparse to dense configurations. Reconstructions were performed at both end systole and end diastole, with EF calculated as the percentage change in cardiac blood volume between these time points. Additionally, global longitudinal strain (GLS) and global circumferential strain (GCS) were derived from the UK Biobank for 15 patients and compared to manual segmentations.

The results in Fig. \ref{fig:clinical_analysis}(a) indicate that the marching cubes method, coupled with smoothing (currently the gold standard in medical image processing), results in significantly different LV reconstructions in sparse images, with a p-value < 0.05, indicating a deviation from the dense fit ground truth. In contrast, GHD+DVS provides more accurate volume quantification even in the sparsest cases, with a p-value of 0.757, as determined by linear regression analysis.

Fig. \ref{fig:clinical_analysis}(b) further illustrates that the marching cubes method tends to underestimate EF when sampling is sparse, leading to values that deviate more significantly from the ground truth. GHD+DVS, however, maintains a stronger correlation with expected EF values, closely aligning with the ground truths.

For both volume and EF quantifications, the marching cubes method with 50 slices was used as the ground truth. In terms of strain analysis, Fig. \ref{fig:clinical_analysis}(c) demonstrates that GHD+DVS closely matches manually calculated strains, underscoring its reliability for clinical application.

%% file: discussion/discussion.tex
\section{Discussion}

To summarize, we propose a novel method, GHD+DVS, for the differentiable reconstruction of the left ventricle myocardial 3D mesh from clinical images, which can be CT, echo, or MRI images. We propose the novel DVS approach to supervise the mesh morphing to fit with image labels. The Winding-DVS is a global loss function that enables easier and better convergence for improved accuracy. We further introduced GHD as a novel mesh morphing approach to representing the 3D mesh, which has the advantages of being naturally smooth without regularization and yet robustly flexible to fix complex shapes. GHD modes are derived from the canonical mesh, which serves as the shape prior, and do not need to be derived from a large dataset of mesh ground truths. This makes it easy to adopt and does not require laborious collection and processing of large data. 
We demonstrate that the GHD+DVS approach has robust performance that challenges the state of the art. In dense data, it performs equally well as the traditional mesh morphing + Chamfer loss approach and outperforms SSM and marching cube approaches. However, with sparse MRI data, existing methods perform poorly, and GHD+DVS is the best-performing approach.

One important utility of GHD+DVS is for the extraction of clinically relevant measurements. We show that it can extract EF, cardiac chamber volumes, and myocardial strains accurately, performing better than the clinical gold standard of marching cubes. In this experiment, we further demonstrated that by applying the approach to different states of dynamically moving organs, such as the heart, we can describe deformations, volume changes, and other dynamic changes, suggesting a wide range of possible applications. Currently, the GHD+DVS is applied frame by frame. However, it can be incorporated into a temporal framework that regularizes for temporal consistency.



\begin{figure}[htbp!]
    \centerline{\includegraphics[width=0.8\columnwidth]{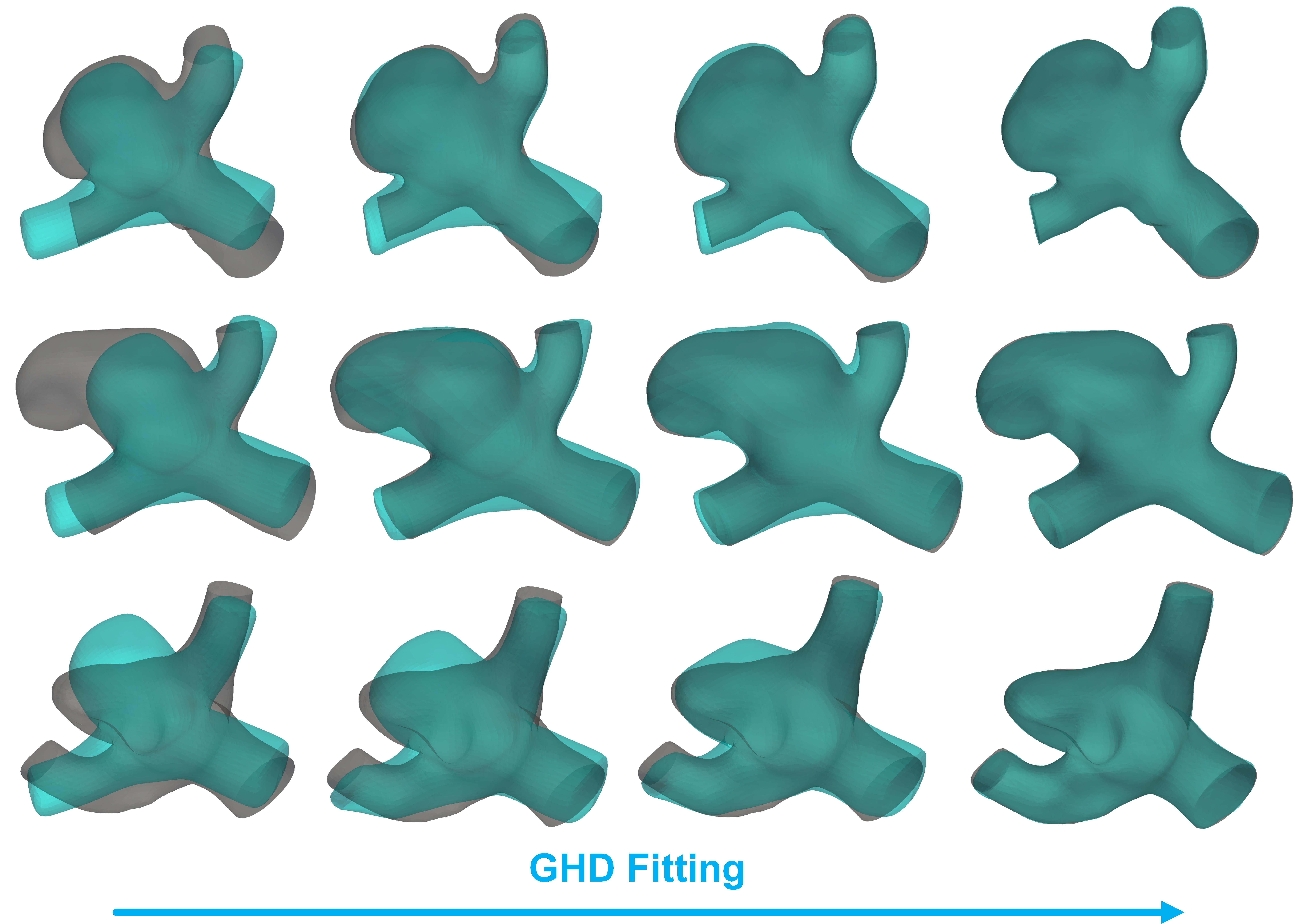}}
    \caption{GHD fitting upon intracranial aneurysms: the first column shows the initial stage of the canonical aneurysm with the target. The following columns show the GHD fitting procession. The last column shows the final GHD fitting results closely matching the target.}
    \label{fig:ghd_aneurysm}
\end{figure}

Besides reconstructing the LV, we propose that the GHD+DVS method can be a universal method for reconstructing various tissues and organs. For example, we successfully applied it in the rebuilding of cranial aneurysms together with the surrounding blood vessels (Fig. \ref{fig:ghd_aneurysm}). The GHD+DVS reconstruction can be good shape inputs for flow dynamics predictor networks, and they may provide better morphological parameters for disease outcome predictions. We further envision that mesh reconstruction can also be applied to the brain, blood vessels, liver, placenta, fetus, limb parts, etc., for various biomedical research and clinical measurements.

Importantly, the GHD-DVS mesh fitting framework is differentiable and can be utilized in deep learning and non-deep learning algorithms requiring mesh reconstruction to specific objectives. For example, it can be used for modeling cardiac biomechanics according to physics constraints while supervised by motions extracted from images, or it can be used for modeling the geometry of the heart while maintaining a concurrent match to cardiac images from different scan modalities (e.g., MRI and echo).

\section{Acknowledgement}

Yihao Luo's PhD project is funded by Imperial College London and the China Scholarship Council (CSC). Dario Sesia's PhD project is funded by an NHLI Endowment studentship and supported by the British Heart Foundation (BHF) DTP (Doctoral Training Programme). This work was supported in part by the ERC IMI (101005122), the H2020 (952172), the MRC (MC/PC/21013), the Royal Society (IEC\textbackslash NSFC\textbackslash211235), the NVIDIA Academic Hardware Grant Program, the SABER project supported by Boehringer Ingelheim Ltd, NIHR Imperial Biomedical Research Centre (RDA01), Wellcome Leap Dynamic Resilience, UKRI guarantee funding for Horizon Europe MSCA Postdoctoral Fellowships (EP/Z002206/1), and the UKRI Future Leaders Fellowship (MR/V023799/1).